\documentclass[11pt,a4paper,english]{article}

\usepackage[utf8]{inputenc}			
\usepackage[T1]{fontenc}				
\usepackage{fancyhdr} 				
\usepackage{amsmath}					
\usepackage{amsfonts}				
\usepackage{amssymb}					
\usepackage{graphicx}				
\usepackage{xcolor}					
\usepackage{multirow} 				
\usepackage{multicol} 				
\usepackage{fancybox}				
\usepackage{caption}					
\usepackage{natbib}					
\usepackage[sc]{titlesec}			
\usepackage[english]{babel}			
\usepackage{wrapfig}

\usepackage[left=2.5cm,right=2.5cm,top=2.5cm,bottom=2.5cm]{geometry}

\begin{document}
\newcommand{\titre}[1]{\noindent \Large \textbf{#1} \normalsize \newline \newline \newline}
\newcommand{\auteurs}[1]{\noindent \large \textbf{#1} \normalsize \newline \newline}
\newcommand{\entite}[3]{${}^{#1}$ \textbf{#2} \newline \textbf{#3} \newline}
\newcommand{\email}[1]{${}^{\ast}$\textbf{#1} \newline}
\newcommand{\lh}{\hbox{\raisebox{0.4em}{\vrule depth 0pt height 1pt width 16cm}}}
\newcommand{\resume}[1]{\lh \noindent \textsc{\textit{RESUME.}} \textit{#1} \newline \newline}
\newcommand{\motsclefs}[1]{\textsc{\textit{MOTS-CL\'EFS.}} \textit{#1} \newline \newline \lh \newline}
\renewcommand{\abstract}[1]{\lh \noindent \textsc{\textit{ABSTRACT.}} \textit{#1} \newline \newline}
\newcommand{\keywords}[1]{\textsc{\textit{KEYWORDS.}} \textit{#1} \newline \newline \lh \newline}
\setcounter{secnumdepth}{4}
\pagestyle{fancy}
\titre{Comparison between purely statistical and multi-agent based approaches for occupant behaviour modeling in buildings}

\auteurs{Khadija Tijani$^{\ast 1,2}$,Ayesha Kashif$^{1,3}$,Quoc Dung Ngo$^{1}$,Stéphane Ploix$^{1}$,Benjamin Haas$^{2}$,Julie Dugdale$^{3}$}
\entite{1}{G-SCOP Sciences pour la Conception l'Optimisation et la Production à Grenoble}{46 avenue Félix Viallet 38031 Grenoble France}
\entite{2}{CSTB Centre Scientifique et Technique du Bâtiment}{84 avenue Jean Jaurès 77420 France}
\entite{3}{LIG Laboratoire d'Informatique de Grenoble}{110 avenue de la chimie Domaine Universitaire de Saint Martin d'hères 38041 Grenoble France}
\email{Lalla-khadija.Tijani@grenoble-inp.fr}

\resume{Ce papier analyse deux approches de modélisation du comportement d'occupants dans le bâtiment. Il compare une approche purement statistique avec une approche basée sur la simulation sociale dans un environnement multi-agent. L'étude concerne les ouvertures de porte dans un bureau.}
\motsclefs{bâtiment, simulation d'occupants, approche statistique, simulation sociale, système multi-agent}

\abstract{This paper analyzes two modeling approaches for occupant behaviour in buildings. It compares a purely statistical approach with a multi-agent social simulation based approach. The study concerns the door openings in an office.}

\section{Introduction and context}
Because of the reduction of energy consumptions, the relative impact of occupants is becoming more important. Thus, the design and the operation of building systems have to take into account the occupant behaviours to improve the ratio between services provided to occupants and the required energy: this is the concept of usage efficiency proposed in \citep{chenailler2012}. Regarding solutions driven by numerical models, occupant behaviours have to be modelled properly as the main function of a building is to provide services to its users. Occupant behaviours have to be considered in the study of building energy efficiency by modelling needs in terms of comfort, energy and health to deduce how these affect the indoor environment. 

Furthermore, Building Energy Management Systems (BEMS) help to optimize energy consumption and also allow occupants to take better decisions regarding energy use. The way it operates involves interactions with the occupants. The BEMS receives signals from the smart grid and gives information about the availability of energy, the price details and energy consumption.  Therefore, the occupants are reactive because they may adapt their behaviours and interact with the BEMS to adapt the proposed energy strategy.

The purpose of this paper is to show that a multi-agent based model can do what a statistical model does and even more by comparing these two approaches: purely statistical approach and multi-agent based design of occupant models. To carry out this comparison, the study focuses on air quality and in particular on occupants' interactions with door openings in an office setting. Such human behaviour depends on several parameters and constraints arising from the external and social environment. A statistical model based on Markov chains is first developed to predict the average state of the door (open or closed) each hour of the day for two months. These results are validated by sensor data. Then, a model of human behaviour focusing on how occupants interact with doors is proposed. The model has been developed using the BRAHMS multi-agent software platform. It is pointed out that a global Markov chain model can be implemented directly in an agent. Then, it is shown that using social simulation with multi-agent systems, the statistic model can be enriched to model interactions between occupants and equipments that can be captured by field study.

\section{State of the art}

The literature suggests that occupant behaviour strongly influences energy consumption patterns and is an important factor for energy waste reduction in buildings \citep{andersen2009}. Models of human behaviour in building simulation tools are usually based on statistical algorithms that predict the probability of an action or event. \citep{dong2009} developed an event based pattern detection algorithm for sensor-based modelling and prediction of user behaviour. They have connected behavioural models based on a Markov model to building energy and comfort management through the EnergyPlus simulation tool for energy calculations. Building simulation tools are based on heat transfer, thermodynamic equations and a human model. The latter is traditionally based on fixed schedules and predefined rule; this does not reflect the actual human behaviour complexity nor reactive and deliberative behaviours. 

In the scientific literature, two different kinds of approaches of occupant behaviour can be found: stochastic approaches, usually based on Markov chains, and multi-agent system approaches.

A first attempt \citep{fritsch1991} to develop a statistical model to predict the state of windows was based on a discrete-time Markov process model to predict transitions between angles of openings. They used Markov chains with six states, each corresponding to a class of opening angles.\citep{page2007} built a time series of presence/absence from the data collected from single person offices and used a Markov chain to reproduce the presence profiles through simulations. \citep{haldi2009} have developed a hybrid stochastic model of window opening, based on three modelling approaches: logistic probability distributions, Markov chains and continuous-time random processes.

The multi-agent approach allows more complex reactive behaviours to be modelled but more parameters must be customized. \citep{abras2010} gave the control of appliances and sources to the software agents that are used in a home automation system. \citep{kashif2014} describes a detailed inhabitant model that represents cognitive, reactive and deliberative behaviours.

\section{Problem statement}
\subsection{Study case}
\subsubsection{The field survey}
In this section, the experiment, which yields data for model design, is presented. Data have been collected from the Grenoble Institute of Technology that contains an engineering school and a research laboratory. The building has been equipped with 165 ENOCEAN sensors on an installation area of 1500 $m^2$, which is divided into three distinct patterns of use:

\begin{itemize}
\item Administration (offices, meeting room, hall, etc)
\item Teaching (classroom, computer room, corridors, etc)
\item Research (offices, meeting room, cafeteria, open-space, etc)
\end{itemize}

This paper focuses on a particular office occupied by researchers because it is equipped with many sensors. An air quality sensor is used for measuring $\rm CO_2$, VOC (Volatile organic compounds), humidity (relative and absolute) and temperature. A presence sensor detects the movement of a person in the room. This type of sensor is useful to validate the presence of a person in the office. This sensor does not return a signal of presence if someone is sitting behind a computer, or if someone is not in the detection area. The contact sensors give the state of doors or windows (open=1/closed=0), the data from these sensors translate the interactions between the occupants and the environment. These sensors provide data using the ENOCEAN radio protocol: information may be lost during communication.

Only the data related to October and November 2013 have been used. All data from the sensors are transmitted to a hub. These data are transmitted at random times, which are different for each sensor. Therefore, each set of data has a variable size, depending on the day and given data. Each sensor provides two vectors, a « time » vector and a « value » vector. Time vector provides the dates in hours that corresponds to the values in the same position in the « value » vector.

\subsection{Study objective}
This paper proposes a common framework to simulate both Markov chains at the group occupant level, and a fine reactive modelling at the individual occupant level with a multi-agent system. The proposed approach focuses firstly on the occupant actions on doors. A group model is designed and simulated and within a multi-agent system using Markov chains. Secondly, the model is enriched by modelling each occupant independently taking into account their interactions with the door. 

\section{Modelling of the group}

\subsection{Purely statistical model}

The Markov chain model is used to predict the door states at each hourly time step. Data processing is carried out to standardize the contact sensor data in setting the data time step to one hour and computing the door states in terms of the ratio of opening time within an hour. Three states of the door are created:

\begin{itemize}
\item door is considered open when the opening duration percentage $\ge$ 80\%
\item door is considered as moving, if the door state changes many times within the considered hour, when 20\% $\le$ opening duration percentage $\le$ 80\%
\item door is considered closed when the opening duration percentage $\le$ 20\%
\end{itemize}

In the data processing, it assumed that the door is closed on weekends and during week days from 8pm to 8am. Two time slots per day have been taken:
\begin{itemize}
\item 1st time slot (working time): 8am $\rightarrow$ 12pm and 2pm $\rightarrow$ 8pm
\item  2nd time slot (lunch time): 12pm $\rightarrow$ 2pm
\end{itemize}

The next step is to build the transition matrix for each time slot. The transitions from one state to another have been counted. Since there are 3 distinct states, the matrix is a 3 * 3 square matrix, which models 9 transitions:

 \begin{eqnarray}
 TM =
 \begin{bmatrix}␣T_{\rm oo}␣&␣T_{\rm om}␣&␣T_{\rm oc}␣\\
T_{\rm mo}␣&␣T_{\rm mm}␣&␣T_{\rm mc}␣\\
T_{\rm co}␣&␣T_{\rm cm}␣&␣T_{\rm cc}␣\end{bmatrix}
\end{eqnarray}

To get the door state for each hour, two steps are required. The first multiplies the transition matrix by the vector representing the previous state. The second one selects a random number at each time step and compare it with the probabilities obtained from the previous step to determine the current state. The vector of door state probabilities is written as:  $P(t) = (P_o(t), P_m(t), P_c(t))$ with $P_o$, the probability to have the state open, $P_m$, the probability to have the state move and $P_c$, the probability to have the state close.

Assume the initial state of the door is closed: $S(0) = (0, 0, 1)$. The next sample time $t=1h$ is computed by: $S(1) = \mathcal{D}\left(S(0) \times TM\right)$ where the function $\mathcal{D}(.)$ stands for selecting a unified random number $r$ and comparing it with each probability element of $P(1)=S(0) \times TM$:
\begin{itemize}
\item if $r \leqslant P_o(1)$ then state is Open 
\item if $P_o(1) \leqslant  r\leqslant P_o(1)+ P_m(1)$ then state is Move
\item if $P_o(1)+P_m(1) \leqslant  r\leqslant P_o(1)+ P_m(1)+ P_c(1)$ then state is Close
\end{itemize}

\begin{figure}
\centering
\includegraphics[scale=0.35]{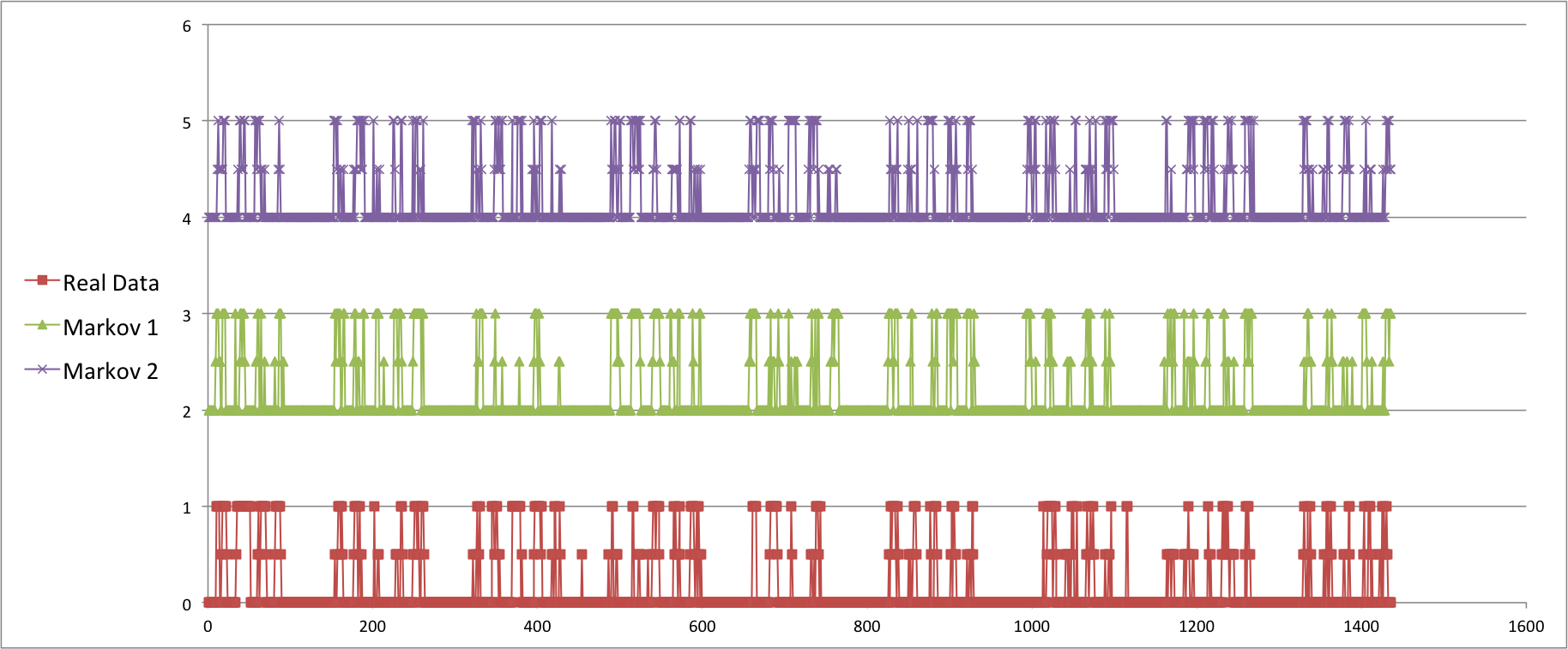}
\caption{Comparison between door states for real data and two Markov simulations}
\label{comparison1}
\end{figure}

Because of random processes, it has been simulated several times to check whether these simulations are close to the recorded data. Figure \ref{comparison1} shows a comparison between the recorded data and two simulation results for the 2 considered months.

\subsection{Multi-agent  system implementation}
In the above section the actions of occupants on the door are simulated using Markov processes. It can also be implemented using a multi agent system approach. The design process consists of agent's perception from the outside environment and internal physical comfort. This perception of the environment are modelled as agents' beliefs. These beliefs lead the agent to go through the cognition phase where the agent makes some desires based on the beliefs. Finally, based on the social and environmental constraints, the desires are transformed into an agent's intention. This finally leads the agent to take some actions on the environment. A change in the environment leads the agent to revise its previous beliefs about the environment and based on the new beliefs repeat the above process.

\begin{figure}
\centering
\includegraphics[scale=0.8]{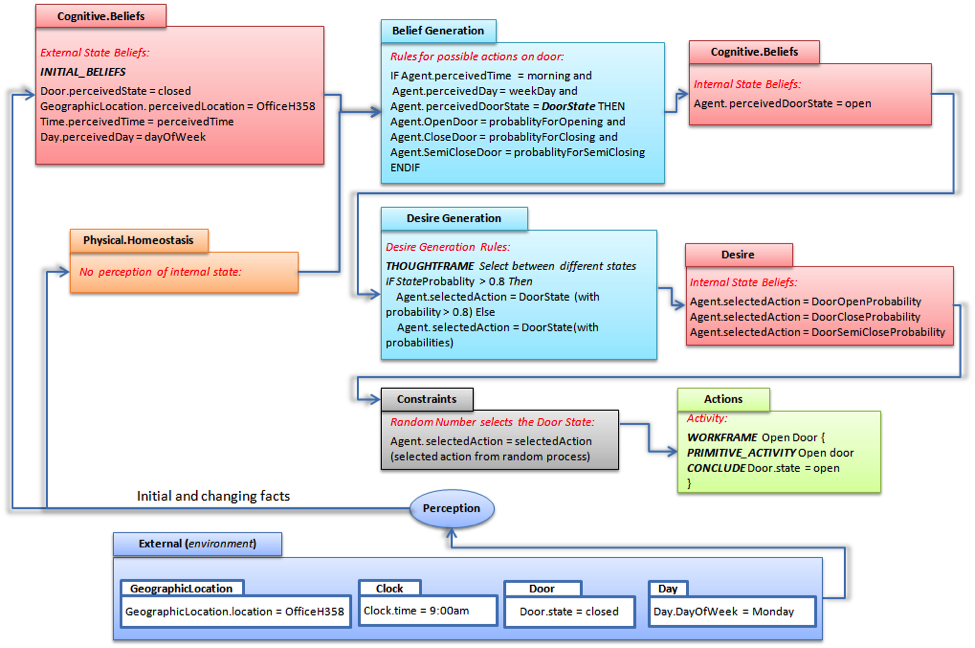}
\caption{Design process to select the next state of the door}
\label{MG1}
\end{figure}

In a first step, all of the occupants in the office are considered as one global group agent, that means that the behaviour of the group agent is actually the behaviour of the Markov process. This group agent perceives some limited information from the outside environment, e.g. the current day and time, the location and the current state of the door which are transformed into agent's beliefs (fig.\ref{MG1}). The door could be OPEN, CLOSE or in a MOVE state. If the current state of the door is open then it has a set of probabilities for the next possible states, that means that the group agent has desires for the door to be in a particular state. Some constraints convert these desires into an intention of the agent to select the next state of the door. This constraint states that the probability for the door to be open at next time step  is higher than 0.8. In case of probabilities between 0.2 and 0.8, a random process decides the selection of the next state of the door. Based on this process, the group agent performs the action on the door and changes its state for the next hour.

\begin{figure}
\centering
\includegraphics[scale=0.8]{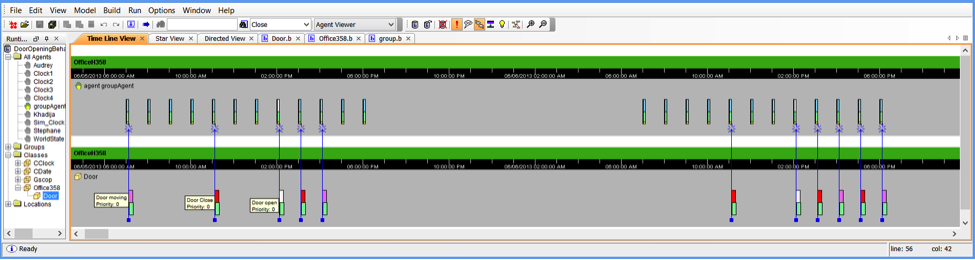}
\caption{The state of the door at each hour due to actions by the group agent}
\label{MG2}
\end{figure}

Figure \ref{MG2} shows the screen shot of simulation results from BRAHMS simulation environment. Only 2 out of 60 days are shown in the figure with actions of the group agent at each hour. The object Door's state when changed is displayed by different colours, e.g. the red colored workframe shows that the door is closed.

From the design process and the simulation results, it can be seen that the next state of the door depends on some simple decision making variables, such as the current state of the door, the time, etc. However, for individual agents that represent actual humans, some additional knowledge, to that of Markov process, is required to predict the future actions on the door. This additional knowledge represents the decisions based on the comfort levels, the influence of other agents around, the social and group behaviours, etc. In the next section a scenario that takes into account more complex behaviour of inhabitants and their resulting actions to change the door state is explained. This shows that not only Markov processes can be implemented using a multi agent approach but also more sophisticated cognitive elements of occupants' behaviours can be introduced in the simulation.

\section{Modelling of individuals: application of a multi-agent based design approach}

\subsection{Field study}

In this section, the office occupants are modelled more precisely. The office is occupied by three persons, Stephane, Khadija and Audrey. Stephane often goes for lectures at the university. Audrey comes to the office one week out of two. Khadija comes to the laboratory every day. In the morning, Khadija usually arrives first, then Stephane and then Audrey. The agent that arrives first opens the door and then either leaves it open or closes it. The three agents have different door opening and closing behaviours during different times of the day. Khadija mostly closes the door after opening it in the morning. However, sometimes, she leaves it open. Audrey always closes the door as she sits very close to it, it disturbs her. Stephane's behaviour is mostly dependent upon the presence of Audrey in the office and he mostly leaves the door open if Audrey is not in the laboratory, otherwise he closes the door if Audrey is present. While Stephane is in the office, visitors sometimes come to see him. Visitors mostly leave the door open while they are in the office and when they leave. Audrey, if present, closes the door after their departure, otherwise it remains open. The agents, however, close the door while going to lunch. In the afternoon, Khadija and Audrey usually go to the cafeteria for a coffee break. Khadija uses to ask Stephane if he wants to go for a coffee. Sometimes he accepts Khadija's proposal but when he is busy, he prefers Khadija to bring him a coffee from cafeteria. If Khadija has to bring coffee for Stephane, and Audrey is not in the university, Khadija leaves the door open as she leaves as she believes that it would be difficult on the way back to open the door with a coffee in each hand. However, if Audrey is in the office and if she is accompanying Khadija, she closes the door before going to cafeteria to get coffee. Stephane sometimes has meetings in a nearby meeting room and he usually leaves the door open when leaving, except if Audrey is in the office: in that case, he closes the door when leaving. Finally, the door is closed at night.

\subsection{Simulation and results}

Each actor is represented by an agent in the environment. This environment monitors the movements of the agent, the activities that the agent performs and its thought processes.

\begin{figure}[h]
\centering
\includegraphics[scale=0.35]{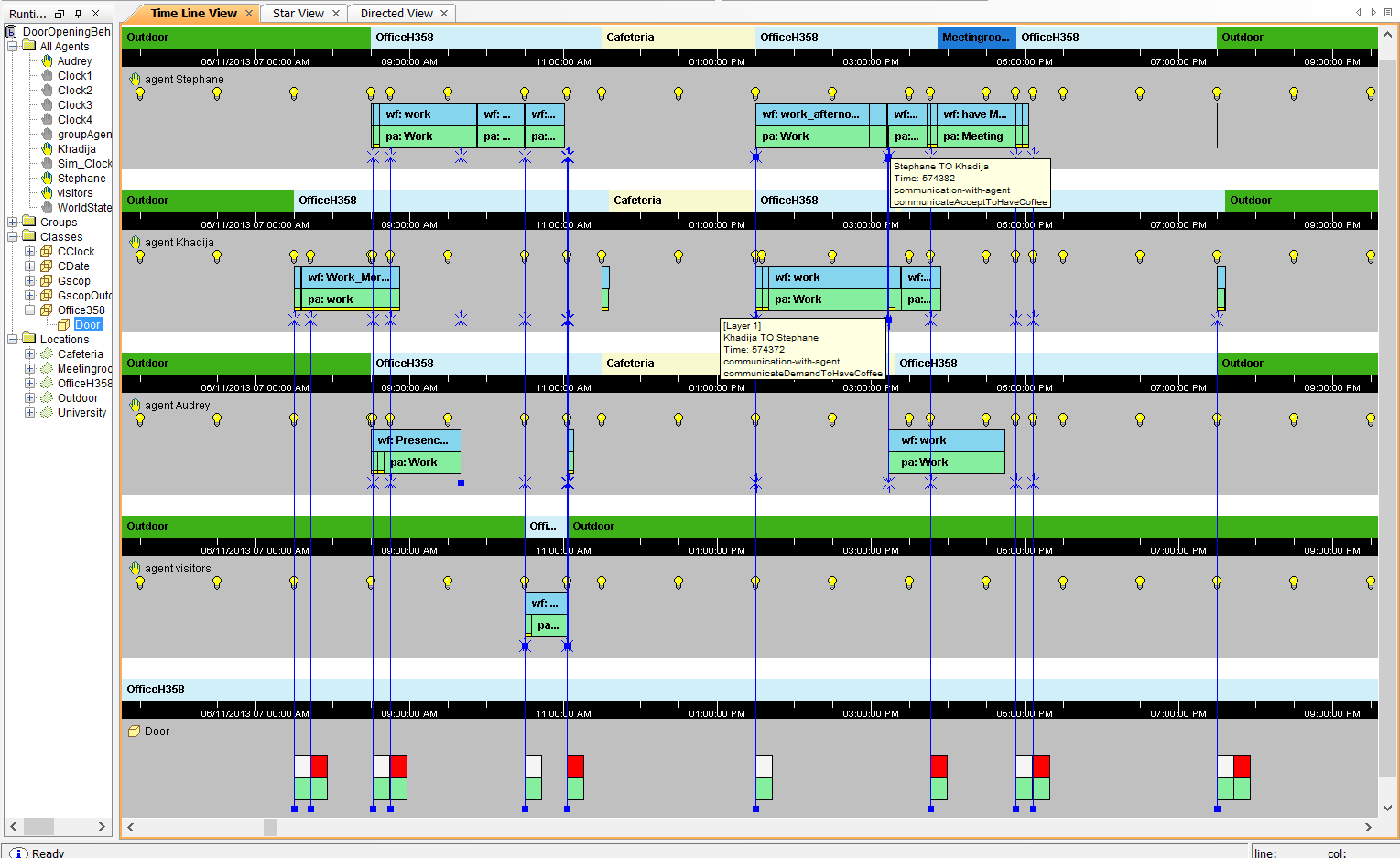}
\caption{Impact of occupants' behaviour on the state of the door in the presence of Audrey agent}
\label{sim6}
\end{figure}

Figure \ref{sim6} shows a screen shot from the simulation where agent Audrey is present in the office. The state of the door in the presence of agent Audrey is kept mostly closed. When the visitors arrive at around 11am, they leave the door open but as soon as they leave, agent Audrey closes the door. Figure \ref{sim6} also shows the communication between the agents  Khadija and Stephane, where agent Khadija asks agent Stephane to have a coffee and  agent Stephane replies. As  agent Audrey is in the office,  agent Stephane closes the door before leaving to the nearby meeting room.

\begin{figure}
 \centering
 \includegraphics[scale=0.45]{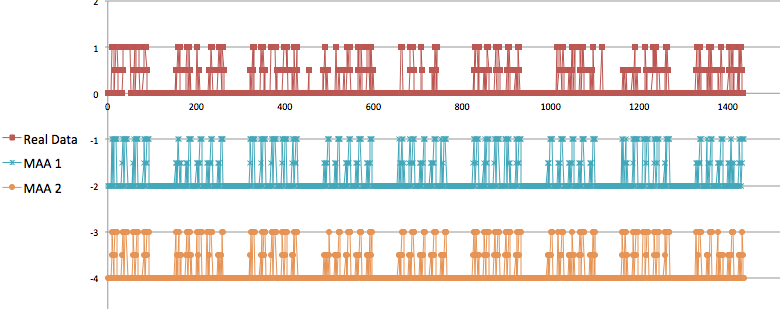}
 \caption{Comparison between the real data and the multi-agent approach model results}
 \label{MAA}
\end{figure}

 Figure \ref{MAA} shows a comparison between the recorded data and the multi-agent approach model results. 
 
\section{Analysis of the results}

The previous sections show that a multi-agent model provides the same results as a purely statistical model. Therefore, theoretically speaking, a multi-agent model could be implemented as a purely statistical model. Multi-agent based social simulation makes it possible to easily implement the behaviours captured during field studies, to model the reasons behind an agent's actions i.e. cognition and reactive behaviours, and to model the interactions between individuals. Figure 6 shows a comparison between the two approaches with the recorded data. This shows that a multi-agent based approach can model the complex behaviours and multiple interactions between occupants.

\begin{figure}
\centering
\includegraphics[scale=0.7]{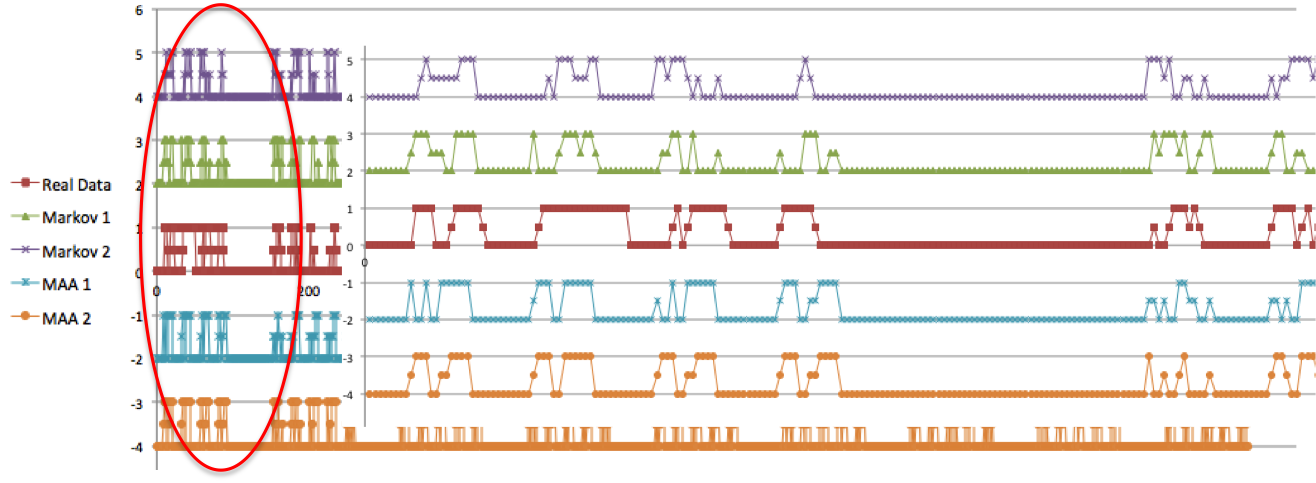}
\caption{Comparison between Multi-Agent Approach results and the recorded data for 2 months}
\label{global}
\end{figure}

\section{Conclusion and future works}

Multi-agent approaches are able to capture the same level of complexity as Markov chain processes. Combined with fields studies, multi-agent approach can propose models which goes beyond Markov chain statistical approaches. However, the complexity of current multi-agent description make it difficult to apply to building energy and indoor air quality simulation. The future work will focus on devising a simplified multi-agent approach for building simulation. A first step will be a connection with $\rm C0_2$ modelling for an integrated evaluation of indoor air quality and actions on doors and windows. 

This work is funded by Agence Nationale de la Recherche project MAEVIA under contract ANR-12-VBDU-0005. Julie Dugdale would like to acknowledge the support of the University of Adger, Norway, to which she is affiliated. 

\bibliographystyle{apalike2}
\bibliography{biblio}

\begin{thebibliography}{}

\bibitem[A.Kashif, 2014]{kashif2014}
A.Kashif (2014).
\newblock {\em Modélisation du comportement humain réactif et délibératif
  avec une approche multi-agent pour la gestion énergétique dans le
  bâtiment}.
\newblock PhD thesis, Université de Grenoble.

\bibitem[C.A.Roulet and P.Cretton, 1991]{fritsch1991}
C.A.Roulet, R.Fritsch, J. and P.Cretton (1991).
\newblock Stochastic model of inhabitant behavior with regard to ventilation.
\newblock {\em Technical report}.

\bibitem[Chenailler, 2012]{chenailler2012}
Chenailler, H. (2012).
\newblock {\em L’efficacité d’usage énergétique : pour une meilleure
  gestion de l’énergie électrique intégrant les occupants dans les
  bâtiments}.
\newblock PhD thesis, Université de Grenoble.

\bibitem[Dong and Andrews, 2009]{dong2009}
Dong, B. and Andrews, B. (2009).
\newblock Sensor based occupancy behaviour pattern recognition for energy and
  comfort management in intelligent buildings.
\newblock {\em 11th Int. Building Perfor- mance Simulation Association (IBPSA)
  Conf. (Glasgow, Scotland; 2009),}, pages 1444--1451.

\bibitem[D.Robinson and F.Haldi, 2009]{haldi2009}
D.Robinson and F.Haldi (2009).
\newblock Interactions with window openings by office occupants.
\newblock {\em Energy and Buildings}, 44:2378--2395.

\bibitem[J.Page, 2007]{page2007}
J.Page, D.Robinson, J. (2007).
\newblock Stochastic simulation of occupant presence and behaviour in
  buildings.
\newblock {\em Proc. Tenth Int. IBPSA Conf : Building Simulation}, pages
  757--764.

\bibitem[R.V.~Andersen and Olesen, 2009]{andersen2009}
R.V.~Andersen, J.~Toftum, K.~A. and Olesen, B. (2009).
\newblock Survey of occupant behaviour and control of indoor environment in
  danish dwellings.
\newblock {\em Energy and Buildings}, 41:11--16.

\bibitem[S.Abras and M.Jacomino, 2010]{abras2010}
S.Abras, S.Ploix, S. and M.Jacomino (2010).
\newblock Advantages of mas for the resolution of a power management problem in
  smart homes, in advances in intelligent and soft computing.
\newblock {\em Springer, Berlin, Heidelberg}, pages 269--278.

\end{thebibliography}

%
%
\end{document}